\documentclass[twocolumn,preprintnumbers,amsmath,amssymb]{revtex4}

\usepackage{graphicx}% Include figure files
\usepackage{dcolumn}% Align table columns on decimal point
\usepackage{bm}% bold math
\usepackage{rotating}% bold math
\usepackage{verbatim}
\usepackage{lipsum}

\begin{document}

\title{Fast Magnetic Reconnection Due to Anisotropic Electron
  Pressure}

\author{P.~A.~Cassak$^1$, R.~N.~Baylor$^1$, R.~L.~Fermo$^1$, 
  M.~T.~Beidler$^1$, M.~A.~Shay$^2$, M.~Swisdak$^3$, J.~F.~Drake$^3$,
  and H.~Karimabadi$^{4,5}$}

\affiliation{$^1$Department of Physics and Astronomy, West Virginia
  University, Morgantown, WV, 26506, USA}

\affiliation{$^2$Department of Physics and Astronomy, University of
  Delaware, Newark, DE, 19716, USA}

\affiliation{$^3$Department of Physics and Institute for Research in
  Electronics and Applied Physics, University of Maryland, College
  Park, MD, 20742, USA}

\affiliation{$^4$Department of Electrical and Computer Engineering,
  University of California at San Diego, La Jolla, CA, 92093, USA}

\affiliation{$^5$SciberQuest, Inc., Del Mar, CA, 92014, USA}

\date{\today}

\preprint{Accepted to {\it Phys.~Plasmas}, February 2015}

\begin{abstract}
  A new regime of fast magnetic reconnection with an out-of-plane
  (guide) magnetic field is reported in which the key role is played
  by an electron pressure anisotropy described by the
  Chew-Goldberger-Low gyrotropic equations of state in the generalized
  Ohm's law, which even dominates the Hall term.  A description of the
  physical cause of this behavior is provided and two-dimensional
  fluid simulations are used to confirm the results.  The electron
  pressure anisotropy causes the out-of-plane magnetic field to
  develop a quadrupole structure of opposite polarity to the Hall
  magnetic field and gives rise to dispersive waves.  In addition to
  being important for understanding what causes reconnection to be
  fast, this mechanism should dominate in plasmas with low plasma beta
  and a high in-plane plasma beta with electron temperature comparable
  to or larger than ion temperature, so it could be relevant in the
  solar wind and some tokamaks.
\end{abstract}

\maketitle

Magnetic reconnection allows for large-scale conversion of magnetic
energy into kinetic energy and heat by changing magnetic topology.  It
occurs in a wide range of environments, such as solar eruptions,
planetary magnetospheres, fusion devices, and astrophysical settings
\cite{Zweibel09}.  One key unsolved problem is what determines the
rate that reconnection proceeds \cite{Cassak12,Karimabadi13}.

In simplified two-dimensional (2D) systems often employed in
simulations, the reconnection rate $E$ is determined by the aspect
ratio of the current sheet, but it is not understood what controls its
length.  In collisional plasmas, current layers are elongated
\cite{Sweet58,Parker57} which make collisional reconnection relatively
slow.  For less collisional 2D systems, elongated layers break and
produce secondary islands
\cite{Matthaeus86,Biskamp86,Loureiro07,Bhattacharjee09}, giving
normalized reconnection rates of 0.01
\cite{Bhattacharjee09,Cassak09a,Huang10}.  However, this is ten times
slower than the fastest rates seen in simulations
\cite{Daughton09,Shepherd10,Daughton12,Ji11,Huang11,Cassak13}.
% To help understand why, there is a desire to understand the
% ``minimal physics'' necessary to produce the fastest rates.

The GEM Challenge \cite{Birn01} showed that the Hall term, when
active, is sufficient to produce short current layers with $E \simeq
0.1$.  The interpretation of this is still under debate
\cite{Karimabadi04,Daughton06,Drake08,Malakit09,Cassak10,Liu14,Tenbarge14}.
One can ask - do other mechanisms limit the length of current layers
that could help explain what causes fast reconnection?

In this Letter, we report that fast reconnection can be caused by
electron pressure anisotropy using the Chew-Goldberger-Low (CGL)
equations of state \cite{Chew56} in the generalized Ohm's law.  This
has not been seen previously because (a) most simulations use no
out-of-plane (guide) magnetic field, but this mechanism requires one
and (b) previous fluid simulations included pressure anisotropies only
in the momentum equation, which does not produce fast reconnection
\cite{Birn01a,Hung11,Meng12}.  This result is distinguished from known
results that off-diagonal elements of the electron pressure tensor
balance the reconnection electric field at the reconnection site
\cite{Vasyliunas75,Hesse99,Hesse06} and agyrotropies contribute near
the reconnection site \cite{Kuznetsova01,Scudder08}.  Neither effect
is present for the CGL equations because the pressure tensor is
gyrotropic.  As with the Hall effect, gyrotropic pressure does not
break the frozen-in condition \cite{Le13}.  Nonetheless, it plays a
crucial role in allowing fast reconnection in this regime.

Gyrotropic pressures are different parallel $p_{||}$ and perpendicular
$p_\perp$ to the magnetic field \cite{Parker58}.  The CGL, or double
adiabatic, equations of state \cite{Chew56} follow rigorously from
kinetic theory in the ideal limit (no heat conduction) with strong
enough magnetic fields so particles are magnetized.  Previous studies
treated gyrotropic pressures in tearing instabilities
\cite{Shi87,Cai09,Chen84,Ambrosiano86,Hesse94,Tanaka11}.
% Pressure anisotropies also affect the shock structure of fast
% reconnection in MHD \cite{Lin94b,Biernat02,Hirabayashi13} and
% kinetic \cite{Karimabadi95,Liu11,Liu11b,Liu12} contexts, arise due
% to particle acceleration in contracting islands
% \cite{Drake06,Schoeffler13}, and prevent the relaxation of
% reconnected islands \cite{Schoeffler11}.
Electron pressure anisotropies have garnered interest lately since
they are self-generated by reconnecting magnetic field lines
\cite{Egedal13}.  The resulting equations of state \cite{Le09} are
valid for guide fields no stronger than the reconnecting magnetic
field.
% It was shown that inclusion of a new equation of state \cite{Le09}
% into fluid models produced current sheets that better agree with
% particle-in-cell simulations \cite{Ohia12}.  Electron anisotropies
% impact the structure of reconnecting current layers
% \cite{Le10b,Ng11,Le13,Le14}.

Simulations are carried out using the two-fluid code F3D \cite{Shay04}
modified to include gyrotropic pressures.  The code updates the
continuity equation, momentum equation, Faraday's law, and pressure
equations.  The electric field ${\bf E}$ is given by the generalized
Ohm's law,
\begin{equation} 
  {\bf E} + \frac{{\bf v} \times {\bf B}}{c} = \frac{{\bf J} \times 
    {\bf B}}{nec} - \frac{1}{ne} \nabla \cdot {\bf p}_{e} + \eta {\bf J} 
  + \frac{m_{e}}{e} \frac{d({\bf J}/ne)}{dt}, \label{genohm}
\end{equation}
where ${\bf v}$ is velocity, ${\bf B}$ is magnetic field, ${\bf J}$ is
current density, $n$ is number density, $e$ is proton charge, ${\bf
  p}_{e}$ is the electron pressure tensor, $\eta$ is resistivity,
$m_e$ is electron mass, and each term on the right can be turned off,
including the Hall term ${\bf J} \times {\bf B}/nec$.
% A non-zero electron inertia term formally necessitates additional
% terms in the momentum equation \cite{REFERENCES}, but these are
% omitted for simplicity.
The momentum equation is
\begin{equation}
  \frac{\partial (\rho {\bf v})}{\partial t} + \nabla \cdot (\rho {\bf v} 
  {\bf v}) = \nabla \cdot \left[ \epsilon \frac{{\bf B} {\bf B}}{4 \pi} - 
    \left( p_{\perp} + \frac{B^2}{8 \pi} \right){\bf I}  \right],
  \label{momeqnaniso}
\end{equation}
where $\epsilon = 1 - 4\pi(p_{||} - p_{\perp}) / B^2$, ${\bf I}$ is
the identity tensor, $\rho$ is mass density, and $p$ is total
(electron plus ion) pressure.  
% In the isotropic limit, the right side becomes $\nabla \cdot \left[
%   {\bf B} {\bf B} / 4 \pi - (p + B^2 / 8 \pi) {\bf I} \right]$.

When pressure anisotropies are used, we employ the CGL equations,
equivalent to $p_{\sigma ||} B^2 / \rho^3$ and $p_{\sigma \perp} /
\rho B$ being constants \cite{Chew56} for species $\sigma$.  We write
them as evolution equations \{see Eqs.~(17) and (18) in
Ref.~\cite{Hesse92}\} with ${\bf E} \cdot {\bf J}$ omitted for
simplicity.
% These equations rigorously follow from the Vlasov equation in the
% limit of an infinitely strong magnetic field.  The pressure
% equations for the Egedal equations of state are {\bf REFERENCES}
%\begin{eqnarray}
%  \frac{\partial p_{||}}{\partial t} + ({\bf v} \cdot \nabla) p_{||} & = & 
%  \left( \frac{2 \pi}{3} \frac{\alpha^{2}+\alpha}{(2\alpha + 1)^2} - 
%    \frac{2n}{(2+\alpha)^{2}} \right) \frac{d \alpha}{dt} + 
%  \frac{2}{2 + \alpha} \frac{dn}{dt} \nonumber \\
%  \frac{\partial p_{\perp}}{\partial t} + ({\bf v} \cdot \nabla) p_{\perp}
%  & = & \left( \frac{nB-n}{(\alpha + 1)^2} \right) \frac{d \alpha}{dt} + 
%  \frac{1}{1 + \alpha} \frac{dn}{dt} + \frac{\alpha}{\alpha + 1} 
%  \frac{d(nB)}{dt}, \nonumber 
%\end{eqnarray}
%where {\bf DEFINITIONS}.
The numerical implementation is benchmarked using Alfv\'en waves and
the firehose and mirror instabilities.  For isotropic plasmas,
$p_\sigma / \rho^{5/3}$ = constant.

All quantities are normalized: magnetic fields to the reconnecting
magnetic field $B_{0}$, densities to the value $\rho_{0}$ far from the
current sheet, velocities to the Alfv\'en speed $c_{A} = B_{0} / (4
\pi \rho_{0})^{1/2}$, lengths to the ion inertial length $d_{i} = c /
\omega_{pi}$, electric fields to $c_{A} B_{0} /c$, resistivities to $4
\pi c_{A} d_{i} / c^{2}$, and pressures to $B_0^2 / 4 \pi$.  The
simulation size is $L_x \times L_y = 204.8 \times 102.4$ in a doubly
periodic domain with 4096 $\times$ 2048 cells.  This system is large
enough that boundaries do not play a role; a steady state prevails for
an extended time.

% To verify the implementation of the anisotropic pressure equations,
% a number of tests are performed in the CGL magnetohydrodynamic (MHD)
% limit.  A shear Alfv\'en wave with $p_{\perp} = 0.25$ and $p_{||} =
% 1$ should have a phase speed of $0.5$.  The firehose instability
% with $p_{\perp} = 0.25$ and $p_{||} = 1.5$ should have a growth rate
% of 2.  The fluid version of the mirror mode instability
% \cite{Kulsrud83} with $p_{\perp} = 1$ and $p_{||} = 0.5$ should have
% a growth rate of 0.2.  Each is consistent with simulation tests.

The initial reconnecting magnetic field is $B_{x}(y) = \tanh[(y +
L_{y}/4) / 0.5] - \tanh[(y - L_{y}/4) / 0.5] - 1$.  Unless otherwise
stated, the guide field is large at $5$ and increases at the current
sheet to balance total pressure.  The density $\rho = 1$ and pressure
$p_{\sigma} = 5$ are initially uniform ($p_{\sigma \perp} = p_{\sigma
  ||} = 5$ for anisotropic).  When electron pressure is evolved, ions
are cold, and vice versa.

All simulations use $m_e = m_i / 25$ unless otherwise stated, which is
acceptable because $E$ is insensitive to $m_e$ \cite{Shay98a} and
length scales for the ions ($c_s / \Omega_{ci} \simeq 0.7$) and
electrons ($d_e = c/\omega_{pe} = 0.2$) are sufficiently separated
\cite{Cassak07a}.
% That the separation of scales is sufficient is confirmed by
% comparing the size of the out-of-plane ion and electron current
% sheets (not shown).
The resistivity is 0.005, chosen so that if reconnection is
Sweet-Parker-like, the layer thickness is $(\eta L_x / 4)^{1/2} \simeq
0.5$ which makes it marginal against secondary islands
\cite{Biskamp86}.  Reconnection is seeded using a coherent magnetic
perturbation of amplitude $0.014$.  Initial random velocity
perturbations of amplitude 0.04 break symmetry.  The equations employ
fourth order diffusion with coefficient $D_{4} = 2.5 \times 10^{-5}$
to damp noise at the grid.
% All simulations are carried out until the system reaches a steady
% nonlinear phase.

\begin{figure}
\includegraphics[width=3.4in]{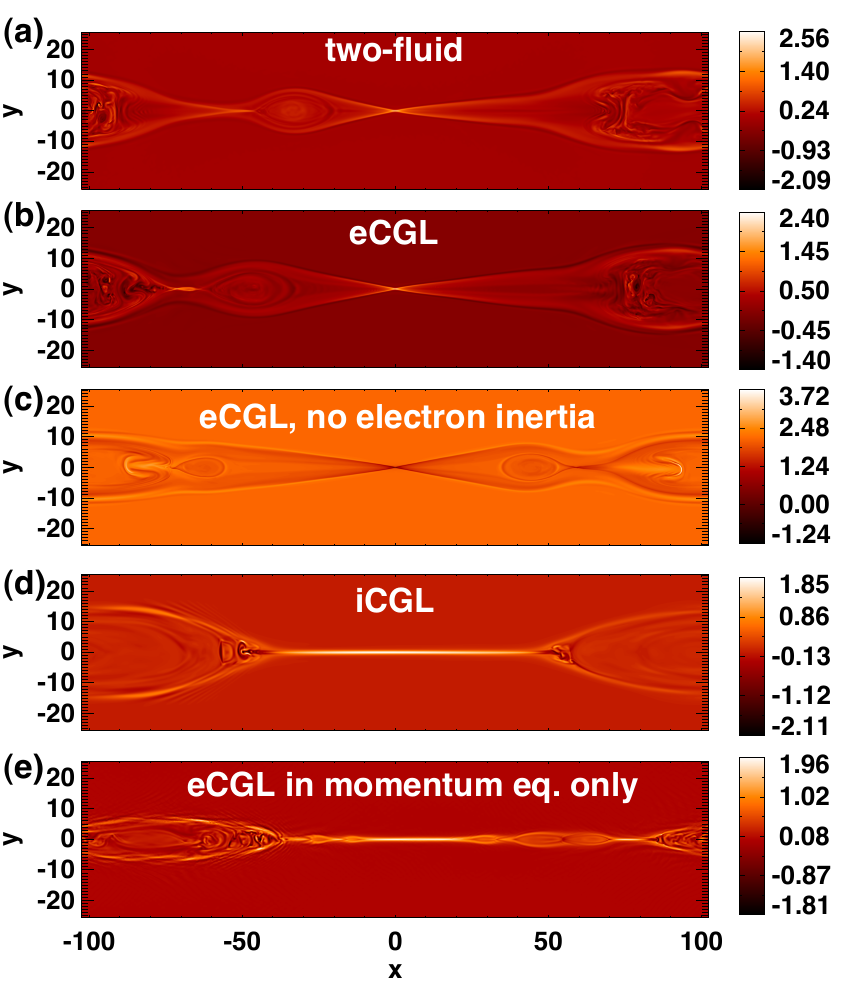}
\caption{\label{fig-currentsheets} (Color) Out-of-plane current
  density $J_z$ using various models: (a) two-fluid, (b) eCGL, (c)
  eCGL with no electron inertia, (d) iCGL, (e) eCGL in the momentum
  equation only.}
\end{figure}

Benchmark simulations
% of known systems are carried out.  In
%resistive-MHD, a standard elongated Sweet-Parker layer appears.  In
using two-fluid simulations (with the Hall term, electron inertia, and
isotropic electron pressure) reveal a well-known open exhaust, as
shown by the out-of-plane current density $J_z$ in
Fig.~\ref{fig-currentsheets}(a).  Various simulations are then
performed without the Hall term.  When the CGL equations are used on
the electrons (which we call eCGL), an open exhaust occurs (panel b).
Panel (c) is for the same system but with $m_e = 0$, showing that
electron inertia does not cause the open exhaust.  Panel (d) is when
the CGL equations are used on the ions (which we call iCGL).  The
current sheet is elongated like in Sweet-Parker reconnection.  To
further identify the key physics, a simulation of an unphysical system
is tested: the electron pressure anisotropy is included in the
momentum equation [Eq.~(\ref{momeqnaniso})] but not in generalized
Ohm's law [Eq.~(\ref{genohm})].  The result is an elongated current
sheet (panel e).  The three with open exhausts are fast, $E \simeq
0.06-0.1$, while the elongated sheets give the Sweet-Parker rate of
0.01.  The thickness of the current sheets in (a) and (b) are near
0.2, showing that layers in eCGL go down to $d_e$ as in two-fluid
reconnection.  In contrast,
%is 0.075 (comparable to the grid scale) for (c), and 
the layer thickness for (d) and (e) is 0.525 and 0.6 (the Sweet-Parker
thickness).  The conclusion is clear: the eCGL equations give rise to
fast reconnection even with no Hall term, and the key physics is the
electron pressure gradient in generalized Ohm's law.

\begin{figure}
\includegraphics[width=3.4in]{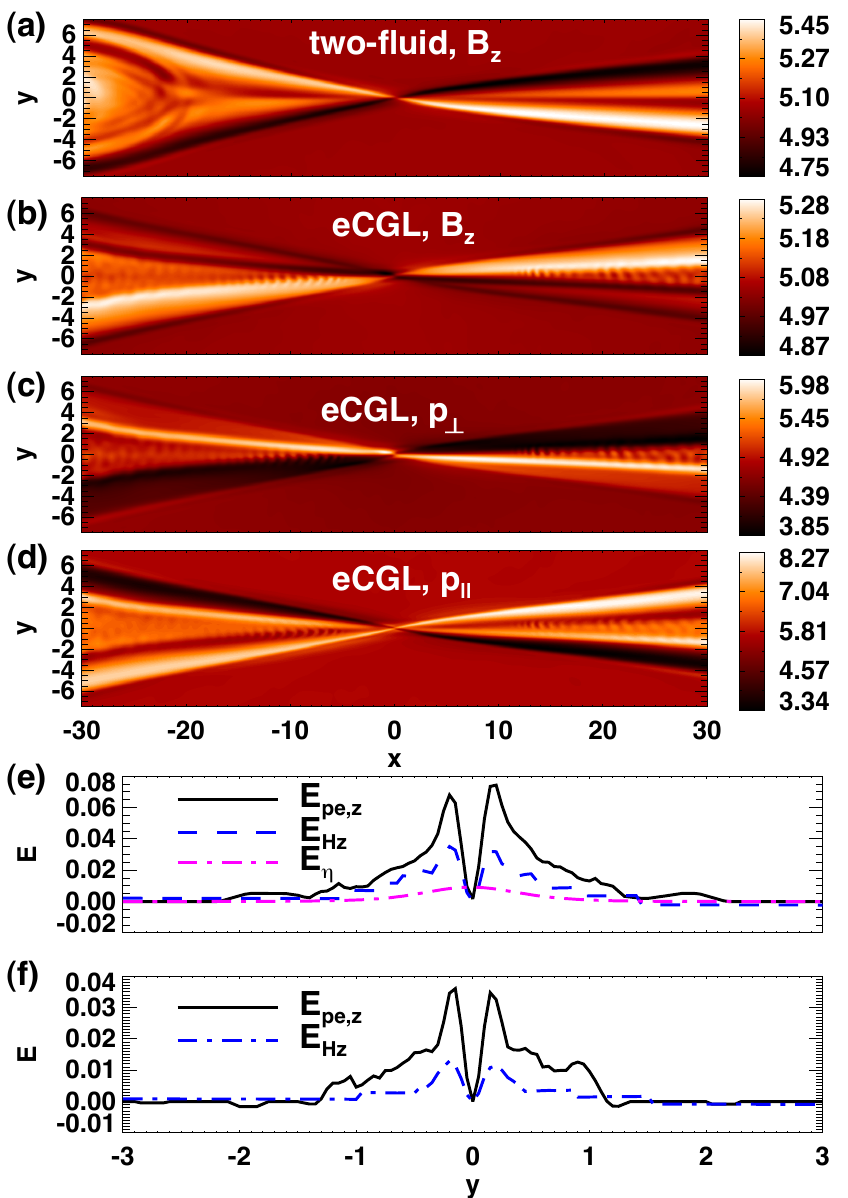}
\caption{\label{fig-quadrupole} (Color) (a) Out-of-plane magnetic
  field $B_z$ in two-fluid reconnection.  For eCGL simulation, (b)
  out-of-plane magnetic field $B_z$, (c) perpendicular electron
  pressure $p_{e \perp}$, and (d) parallel electron pressure
  $p_{e||}$. Reconnection electric field contributions in the (e)
  eCGL, two-fluid, and iCGL simulations and (f) a simulation with eCGL
  and the Hall term.}
\end{figure}

The physics when electron pressure anisotropies dominate bears
similarities to Hall reconnection with a guide field \cite{Kleva95}.
The $z$ component of Eq.~(\ref{genohm}) in terms of the flux function
$\psi$ defined as $E_z = (1/c) \partial \psi / \partial t$ is
\begin{equation}
  \frac{\partial \psi}{\partial t} + \left({\bf v} - \frac{{\bf J}}{ne}
  \right) \cdot \nabla \psi = - \frac{c}{ne} (\nabla
  \cdot {\bf p}_{e})_z + \left(\frac{\eta c^2}{4 \pi} +
    \frac{d}{dt} d_e^2 \right) \nabla^2 \psi. \label{flux}
\end{equation}
With the Hall term present, the left side reveals magnetic flux is
convected by electrons, so electrons carrying the current drag the
reconnecting magnetic field out of the plane \cite{Mandt94}.  This
produces a quadrupole out-of-plane magnetic field $B_z$
\cite{Sonnerup79}, shown for the two-fluid simulation in
Fig.~\ref{fig-quadrupole}(a).  With a strong guide field, the gas
pressure (not shown) develops a quadrupole with opposite polarity to
maintain total pressure balance \cite{Kleva95}.

Without the Hall term, Eq.~(\ref{flux}) implies magnetic flux is
convected by ions \cite{Karimabadi04b}.  The magnetic field is dragged
out of the plane by ions, giving a $B_z$ quadrupole with opposite
polarity as in Hall reconnection, displayed for the eCGL simulation in
Fig.~\ref{fig-quadrupole}(b).  (An instability is visible in the
exhaust.  The system is not firehose or mirror unstable; it is likely
a drift instability.)  To balance total pressure, $p_{e\perp}$
develops a quadrupole of opposite polarity, displayed in
Fig.~\ref{fig-quadrupole}(c).  The density (not shown) develops a
quadrupole like that of $p_{e\perp}$.  This requires a parallel
electric field pointing from low density to high, which comes from a
parallel electron pressure with high $p_{e||}$ in regions of low
$p_{e\perp}$ so $p_{e||}$ has a quadrupole of opposite polarity as
$p_{e\perp}$, shown in Fig.~\ref{fig-quadrupole}(d).  Thus, an
electron pressure anisotropy is self-generated.  It contributes to the
reconnection electric field as $E_{pe,z} = -(1/ne) (\nabla \cdot {\bf
  p}_e)_z = -(1/ne) ({\bf B} \cdot \nabla)[(p_{e\perp} - p_{e||}) B_z
/ B^2]$, plotted as the solid line in a vertical cut through the
X-line in Fig.~\ref{fig-quadrupole}(e).
% This term is the major contributor to the reconnection electric
% field between electron and ion scales.
For comparison, the dashed line shows the Hall electric field $E_{Hz}$
in the two-fluid simulation and the dash-dot line shows the resistive
electric field $E_{\eta}$ in the iCGL simulation.  The structure of
$E_{pe,z}$ is similar to the known $E_{Hz}$ profile in two-fluid
reconnection.  Note that eCGL with $m_e = 0$ also has quadrupoles, but
iCGL with slow reconnection does not.

The guide field is key to the physics.  If it is too large, the ion
Larmor radius falls below electron or resistive scales which prevents
fast reconnection, analogous to Hall reconnection \cite{Aydemir92}.
If it is too small, the pressure change due to the $B_z$ quadrupole is
small, so the effect in the previous paragraph is negligible.  We
quantify this by finding when $E_{pe,z}$ is dominated by other
contributions to Ohm's law, which for the present simulations is the
resistive term.  A scaling analysis gives $E_{pe,z} / E_\eta \sim 0.1
\beta_{e0} (B_g / B_{rec}) / (2 \eta c^2 / 4 \pi c_A d_i)$, where 0.1
is $E$ for fast reconnection, $\beta_{e0} = 8 \pi p_{e0} / (B_{rec}^2
+ B_g^2)$ is the electron plasma beta, and $B_{rec}$ and $B_g$ are the
reconnecting and guide field strengths.  This ratio is small if $B_g$
is sufficiently big or small.
%Another test comes from finding the parameter regime when dispersive
%waves are present and seeing if reconnection is fast for that regime.
%For the anisotropic whistler-like wave, it dominates the shear
%Alfv\'en wave when $-k_{||}^2 k_{\perp}^2 c_A^2 d_i^2
%[\partial(p_{e||} - p_{e\perp}) / \partial p_B]_0 \gg k_{||}^2 c_A^2$.
%For eCGL, the derivative is $-(3/2) \beta_0$, so the condition for the
%dispersive wave to be important becomes $(3/2) \beta_{0} k_{\perp}^2
%d_i^2 \gg 1$.  {\bf GOT TO HERE!!}  In the dispersive wave picture,
%$k_\perp$ corresponds to $(1 / \delta) B_{g}/(B_{rec}^2+B_g^2)^{1/2}$,
%where $\delta$ is the thickness of the current layer, $B_{rec}$ is the
%reconnecting magnetic field and $B_g$ is the guide field.
%Collisionless reconnection is likely only fast if $\delta$ is greater
%than $d_e$, as is the case for Hall reconnection \cite{Aydemir92}.
%Using $\beta_0 = 8 \pi p_0 / (B_{rec}^2 + B_g^2)$, the condition
%becomes $4 \pi (3 p_0) (m_i / m_e) \gg (B_{rec}^2 + B_g^2)^2 / B_g^2$.
%This condition is two constraints when written in terms of the guide
%field strength.  For large guide field, the condition is $B_g^2 \ll 4
%\pi (3 p_0) (m_i / m_e)$, which physically corresponds to the ion
%scale falling below the electron scale for large guide field.  For
%small guide field, the condition is $B_g^2 \gg B_{rec}^4 (m_e / m_i) /
%4 \pi (3 p_0)$, which physically corresponds to the limit where
%$k_\perp$ is small in the reconnection plane.
We confirm this in simulations with $p_{e\perp} = p_{e||} = 1$; the
predicted range is $0.05 \ll B_g \ll 20$.
% For these parameters, the predicted range of guide fields which
% permit fast reconnection is $0.12 \ll B_g \ll 8.66$.
(Formally, the CGL model is invalid for small $B_g$, so this tests
fundamental physics independent of the appropriateness of the CGL
model.)  We find reconnection is Sweet-Parker-like for $B_g = 0$ and
0.1, has a short current layer with $E \simeq 0.03$ for a transitional
guide field $B_g = 0.25$, is fast with $E \simeq 0.05$ for $B_g = 0.5,
5,$ and 7.5, and is again Sweet-Parker-like for $B_g = 15$.  These
results agree with the prediction.

This system yields an interesting way to study the cause of fast
reconnection.  It was proposed that reconnection is fast if linear
perturbations to a homogeneous equilibrium permit dispersive waves
with faster phase speeds at smaller scales \cite{Rogers01}, such as
the whistler or kinetic Alfv\'en wave in Hall-MHD.  This has been
controversial because reconnecting fields are not homogeneous.  
% The present system offers a new regime to make further tests.

We present the linear theory of a plasma with pressure anisotropies,
the Hall term, and electron inertia.
% arbitrary pressures is obtained from Eqs.~(\ref{genohm}),
% (\ref{momeqnaniso}), Faraday's law and the continuity equation, with
% the Hall term included for completeness and allowing for equilibrium
% anisotropies.
Rather than using CGL, we generalize by taking ion and electron
pressures to be arbitrary functions of $\rho$ and $B$, {\it i.e.},
$p_{\sigma \perp} = p_{\sigma \perp}(\rho,B)$ and $p_{\sigma ||} =
p_{\sigma ||}(\rho,B)$.  This captures adiabatic, CGL, and Egedal's
equations of state \cite{Le09,Egedal13}.
% or any other closure model that can be parametrized this way.
The dispersion relation relating the frequency $\omega$ to the
wavevector ${\bf k}$ is
\begin{equation}
\omega^6 - C_4 \omega^4 + C_2 \omega^2 - C_0 = 0, \label{disprel}
\end{equation}
where
\begin{widetext}
\begin{eqnarray}
  C_4 & = & \frac{k_{\perp}^2 c_A^2}{D} \left[1 + 
    \left(\frac{\partial p_{\perp}}{\partial p_{B}} \right)_0 \right]
  + \frac{2 \tilde{\epsilon} k_{||}^2 c_A^2}{D} + \left(\frac{\partial (k_{||}^2 p_{||}
      + k_{\perp}^2 p_{\perp})}{\partial \rho} \right)_0 
  + \frac{\tilde{\epsilon}_e k_{||}^2 c_A^2 d_i^2}{D^2} \left\{ k^2 -
    k_{\perp}^2 \left[ 2 \tilde{\epsilon}_e - 2 + \left( 
        \frac{\partial (p_{e||} - p_{e\perp})}{\partial p_B} \right)_0 \right]  
  \right\} \nonumber \\
  C_2 & = & \frac{\tilde{\epsilon}^2 k_{||}^4 c_A^4}{D^2} + \frac{\tilde{\epsilon} k_{||}^2 
    k_{\perp}^2 c_A^4}{D^2} \left[ 1 + \left( \frac{\partial p_{\perp}}{\partial 
        p_B} \right)_0 \right] 
  + \frac{k_{||}^2 c_A^2 (k_{\perp}^2 + 2\tilde{\epsilon} k_{||}^2)}{D} \left( 
    \frac{\partial p_{||}}{\partial \rho} \right)_0 
  + \frac{k_{||}^2 k_{\perp}^2 c_A^2}{D} \left[ \left( \frac{\partial p_{\perp}}
      {\partial \rho} \right)_0
    + \left( 
      \frac{\partial (p_{||},p_{\perp})}{\partial(\rho, p_B)}\right)_0 
  \right] 
  \nonumber \\
  & & +
  \frac{\tilde{\epsilon}^2 k_{||}^2 k^2 c_A^2 d_{i}^2}{D^2} \left[ k_{||}^2 \left( 
      \frac{\partial p_{||}}{\partial \rho}\right)_0 + k_{\perp}^2 \left( 
      \frac{\partial p_{\perp}}{\partial \rho}\right)_0\right]
  + \frac{\tilde{\epsilon}_e k_{||}^2 k_{\perp}^2 c_A^2 d_i^2}{D^2} [k_{||}^2 
  (2 \tilde{\epsilon} - 1) + k_{\perp}^2 ]
  \left(\frac{\partial (p_{e||}-p_{e\perp})}{\partial \rho} \right)_0 
  \nonumber \\
  & & 
  + \frac{\tilde{\epsilon}_e k_{||}^2 k_{\perp}^2 c_A^2 d_i^2}{D^2} \left[ (2 \tilde{\epsilon}_e 
    - 2) \left(\frac{\partial ( k_{||}^2 p_{||} + k_{\perp}^2 p_{\perp})}
      {\partial \rho} \right)_0 
    -   \left( \frac{\partial(k_{||}^2 p_{i||} + k_{\perp}^2 p_{i\perp},p_{e||}-
        p_{e\perp})}{\partial(\rho,p_B)} \right)_0 - k^2 \left( \frac{\partial(
        p_{e\perp},p_{e||})}{\partial(\rho,p_B)} \right)_0 \right]
  \nonumber \\
  C_0 & = & \frac{\tilde{\epsilon} k_{||}^4 c_A^4}{D^2} \left[ (\tilde{\epsilon} k_{||}^2 
    + k_{\perp}^2) \left( \frac{\partial p_{||}}{\partial \rho} \right)_0 
    + k_{\perp}^2  (1 - \tilde{\epsilon}) \left( 
      \frac{\partial p_{\perp}}{\partial \rho} \right)_0  
    + k_{\perp}^2  \left( \frac{\partial(p_{||},p_{\perp})}
      {\partial(\rho,p_B)} \right)_0  \right]. \nonumber 
\end{eqnarray}
\end{widetext}
Here, $D = 1 + k^2 d_e^2$, $p_B = B^2 / 8 \pi$ is the magnetic
pressure, $\tilde{\epsilon} = 1 - 4 \pi (p_{||0}-p_{\perp 0}) / B_0^2$ is the
equilibrium anisotropy parameter for the total pressure, $\tilde{\epsilon}_e$
is similarly defined for the electrons, $\partial(A_1,A_2)
/ \partial(x,y) = (\partial A_1 / \partial x)(\partial A_2 / \partial
y) - (\partial A_1 / \partial y)(\partial A_2 / \partial x)$ is a
Poisson bracket-type operator, and the $0$ subscript denotes
equilibrium quantities.  This reduces to known results in the limits
of anisotropic-MHD with the CGL equations ($d_\sigma \rightarrow 0,
p_{\sigma\perp} / \rho B = $ constant, $p_{\sigma ||} B^2 / \rho^3$ =
constant) \cite{Hau07} and isotropic two-fluid ($p_{\sigma ||} =
p_{\sigma \perp}$) \cite{Rogers01}.

We find pressure anisotropies introduce dispersive waves even when the
Hall term is absent.  All terms in Eq.~(\ref{disprel}) with $d_i^2$
give dispersive waves.  For the high $\omega$, high $k$ with $\tilde{\epsilon}
= \tilde{\epsilon}_{e} = 1$ limit, $\omega^2 \simeq C_4 \simeq (k_{||}^2 c_A^2
d_i^2 / D^2) \{ k^2 - k_{\perp}^2 [\partial (p_{e||} - p_{e\perp})
/ \partial p_B]_0 \}$.  The $k^2$ term comes from the Hall term and is
the standard whistler wave, while the $k_{\perp}^2$ term is a
whistler-like wave coming from the electron pressure anisotropy in
generalized Ohm's law.  For eCGL, this is $\omega^2 = (3/2) \beta_{0}
k_\parallel^2 k_\perp^2 c_A^2 d_i^2$, where $\beta_0 = p_0 / p_B$.
Similarly, following Ref.~\cite{Rogers01}, the intermediate frequency
range gives $\omega^2 \simeq C_2 / C_4$.  In the $k_{\perp} \gg
k_{||}$, short wavelength, cold ion limit, this yields $\omega^2 =
(k_{||}^2 k^2 c_A^2 d_{i}^2 / D^2) \{ \left( \partial p_{e||}
  / \partial \rho \right)_0 - [ \partial( p_{e\perp},p_{e||})
/ \partial(\rho,p_B) ]_0 \} / [c_A^2 / D + (c_A^2 / D) (\partial
p_{e\perp} / \partial p_{B})_0 + (\partial p_{e\perp} / \partial
\rho)_0]$.  In the low $\beta_0$ limit, the first term gives the
standard kinetic Alfv\'en wave, while the second is a kinetic
Alfv\'en-type wave arising from the pressure anisotropy.  In eCGL,
this wave has $\omega^2 = (5/2) \beta_0 k_\parallel^2 k^2 d_i^2
(p_0/\rho_0)$.

There are many ways to test the dispersive wave model.  For cold ions,
dispersive waves from anisotropies persist.  However, they vanish
identically for cold electrons.  The dispersive wave model predicts
fast reconnection for eCGL but slow reconnection for iCGL, consistent
with our simulations.  Eq.~(\ref{disprel}) implies there are
dispersive waves without the Hall term when there is an equilibrium
pressure anisotropy, independent of the equations of state, consistent
with previous studies \cite{Ambrosiano86,Guo03}.

Interestingly, when Egedal's equations of state \cite{Le09,Egedal13}
are employed in simulations without the Hall term, reconnection is
Sweet-Parker-like (J.~Egedal, private communication).  Thus, simply
having an electron pressure anisotropy is insufficient to cause fast
reconnection; the pressure anisotropy must have a particular form.
Fluid modeling of other equations of state could provide insight about
what physically sets the length of the current layer.
% It is also important to develop a better understanding of the system
% parameters for which the CGL and other closure models are applicable
% in physical systems.

We now show that electron pressure anisotropies can dominate the Hall
term in real systems.  First, we have performed particle-in-cell
simulations with parameters similar to the fluid simulations,
confirming that the CGL model ($p_{||} \propto \rho^3 / B^2$ and
$p_{\perp} \propto \rho B$) is reasonably reproduced (plots not
shown).  Also, electron pressure anisotropies dominate the Hall term
for the parameters of the simulations in Fig.~\ref{fig-currentsheets}.
Fig.~\ref{fig-quadrupole}(f) shows a simulation of eCGL with the Hall
term; the contribution to the reconnection electric field of the
pressure anisotropy (solid line) dominates the Hall term (dashed
line).

We suspect electron pressure anisotropy dominates when dispersive wave
terms due to the anisotropy dominate the standard whistler and kinetic
Alfv\'en waves in Eq.~(\ref{disprel}).  In $C_2$, the first term with
$d_i^2$ gives the standard kinetic Alfv\'en wave.  The second term
with $d_i^2$ is the most important term arising from the electron
pressure anisotropy (by a factor of $\beta$, which is small for many
systems of interest).  In the $\tilde{\epsilon}_e = \tilde{\epsilon} =
1$ limit, a simple calculation reveals that the electron pressure
contribution of the kinetic Alfven wave is completely cancelled by
part of the electron pressure anisotropy.  This implies that it {\it
  always} dominates the Hall term when $T_e > T_i$ with low $\beta$.
Therefore, when $T_e > T_i$, the anisotropy is the dominant mechanism
for the entire parameter regime previously thought to be the kinetic
Alfv\'en regime of reconnection \cite{Rogers01} - low $\beta$, high
in-plane $\beta$ based on $B_{rec}$, and strong guide field (but not
strong enough to make the ion Larmor radius smaller than $d_e$).
Physically, $T_i$ needs to be smaller than $T_e$ because if it is
large enough, it can dominate the electron pressure effect discussed
here.

There are physical systems where reconnection in this parameter regime
could occur.  The solar wind and some tokamaks are low $\beta$ where
significant guide fields are expected and $T_e > T_i$ is possible.

% The present results may be relevant for explaining a recent study
% \cite{Liu14} which found fast reconnection in the large guide field
% limit in a parameter regime in which the Rogers et al.~study
% \cite{Rogers01} would have predicted no fast reconnection.  The
% present results suggest it is possible to get dispersive waves even
% in the absence of the Hall term.  It would be interesting to revisit
% this study to investigate the spectrum of modes in this regime.

We gratefully acknowledge support by NSF grant AGS-0953463 (PAC) and
NASA grants NNX10AN08A (PAC), NNX14AC78G (JFD), NNX14AF42G (JFD), and
NNX11AD69G (MAS).  This research used computational resources at the
National Energy Research Scientific Computing Center and NASA Advanced
Supercomputing.  We thank M.~Kuznetsova, J.~Egedal, and C.~Salem for
helpful conversations.

%\bibliographystyle{apsrev}
%\bibliography{bib}

\end{document}